\documentclass[sigconf, authorversion]{acmart}

\newcommand\pr[1]{P#1}

\AtBeginDocument{%
  }

\copyrightyear{2025}
\acmYear{2025}
\setcopyright{cc}
\setcctype{by}
\acmConference[ITiCSE 2025]{Proceedings of the 30th ACM Conference on Innovation and Technology in Computer Science Education V. 1}{June 27-July 2, 2025}{Nijmegen, Netherlands}
\acmBooktitle{Proceedings of the 30th ACM Conference on Innovation and Technology in Computer Science Education V. 1 (ITiCSE 2025), June 27-July 2, 2025, Nijmegen, Netherlands}\acmDOI{10.1145/3724363.3729093}
\acmISBN{979-8-4007-1567-9/2025/06}

\makeatletter
\gdef\@copyrightpermission{
  \begin{minipage}{0.3\columnwidth}
   \href{https://creativecommons.org/licenses/by/4.0/}{\includegraphics[width=0.90\textwidth]{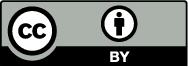}}
  \end{minipage}\hfill
  \begin{minipage}{0.7\columnwidth}
   \href{https://creativecommons.org/licenses/by/4.0/}{This work is licensed under a Creative Commons Attribution International 4.0 License.}
  \end{minipage}
  \vspace{5pt}
}
\makeatother

\usepackage{subfigure}

\begin{document}

\title[Probing the Unknown]{Probing the Unknown: Exploring Student Interactions with Probeable Problems at Scale in Introductory Programming}

\author{Paul Denny}
\orcid{0000-0002-5150-9806}
\affiliation{
  \institution{University of Auckland}
  \city{Auckland}
  \country{New Zealand}
}
\email{paul@cs.auckland.ac.nz}

\author{Viraj Kumar}
\orcid{0000-0002-2252-0141}
\affiliation{%
  \institution{Indian Institute of Science}
  \city{Bengaluru}
  \country{India}
}
\email{viraj@iisc.ac.in}

\author{Stephen MacNeil}
\orcid{0000-0003-2781-6619}
\affiliation{%
\institution{Temple University}
\city{Philadelphia}  
\state{PA}  
\country{USA}
}  
\email{stephen.macneil@temple.edu}

\author{James Prather}
\orcid{0000-0003-2807-6042}
\affiliation{
  \institution{Abilene Christian University}
  \city{Abilene, TX}
  \country{USA}
}
\email{james.prather@acu.edu}

\author{Juho Leinonen}
\orcid{0000-0001-6829-9449}
\affiliation{  
\institution{Aalto University}
\city{Espoo}  
\country{Finland}  
}  
\email{juho.2.leinonen@aalto.fi}


\begin{abstract}

Introductory programming courses often rely on small code-writing exercises that have clearly specified problem statements.  This limits opportunities for students to practice how to clarify ambiguous requirements -- a critical skill in real-world programming.  In addition, the emerging capabilities of large language models (LLMs) to produce code from well-defined specifications may harm student engagement with traditional programming exercises.  This study explores the use of ``Probeable Problems'', automatically gradable tasks that have deliberately vague or incomplete specifications.  Such problems require students to submit test inputs, or `probes', to clarify requirements before implementation.  Through analysis of over 40,000 probes in an introductory course, we identify patterns linking probing behaviors to task success. 
Systematic strategies, such as thoroughly exploring expected behavior before coding, resulted in fewer incorrect code submissions and correlated with course success.  
Feedback from nearly 1,000 participants highlighted the challenges and real-world relevance of these tasks, 
as well as benefits to critical thinking and metacognitive skills. 
Probeable Problems are easy to set up and deploy at scale, and help students recognize and resolve uncertainties in programming problems.
\end{abstract}



\maketitle

\section{Introduction}

In real-world programming contexts, requirements are often conveyed in natural language, which is prone to ambiguity, misinterpretation, and the need for clarification \cite{shah2015resolving, gervasi2019ambiguity}. Failure to clarify ambiguities early can be costly in software development~\cite{fernandez2017naming}.  Thus, identifying gaps and clarifying missing details in specifications are essential practices for developers.  Despite their importance, they are often not emphasized in introductory programming courses, where the focus tends to be on solving well-defined problems with clear, detailed specifications written by the instructor \cite{allen2018weekly, broeders2023improving, decker2024transforming}.  This approach has, at least in part, been popularized by the use of automated grading tools that can instantly provide feedback on student code but are typically used with well-defined problems \cite{messer2024automated, gramoli2016mining, baniassad2021stop}.

While automated grading can scale well and is widely used in programming courses, the arrival of large language models (LLMs) has presented a significant challenge.  Typical introductory-level programming tasks can now be reliably solved by LLMs \cite{prather2023robots, denny2024cacm}.  This means that students can simply provide a programming problem statement -- exactly as defined by their instructor -- to an LLM as a prompt to generate working code. This not only reduces student engagement with programming tasks, but it may lead students to question the value of writing code for well-specified tasks.

One approach to address this challenge is the use of programming tasks with \emph{deliberately vague or incomplete specifications}.  Unlike traditional tasks with well-defined requirements, such problems are resistant to trivial solutions from LLMs because they lack the details required for generating a correct response.  More importantly, these tasks provide students with an opportunity to practice identifying gaps and ambiguities, and to pose questions to fill in the missing pieces.  As an example, consider the request to \emph{``search an array for the smallest even value''}.  While this initially seems straightforward, several key clarifications are needed in order to implement a function to solve this task:  

\begin{itemize}
    \item Should the function return a value, or print output directly?
    \item What should be calculated, an index position or a value?
    \item What should happen if there are no even values present?
    \item What should happen if the smallest even value occurs at multiple positions within the array?
    \item If multiple positions are indeed wanted, then in what order?
\end{itemize}


In this paper, we explore how students approach and perceive the challenge of seeking answers to such clarifying questions.

\begin{table}[ht]
\centering
\caption{Possible ambiguities in the problem statement \emph{``find the first vowel in a string''}, each with a corresponding `probe' (test input) and  `clarification' (expected output).}
\begin{tabular}{@{}l|l|l@{}}
\toprule
\textbf{Ambiguity} & \textbf{Probe} & \textbf{Clarification} \\
\midrule
Return character or index? & \texttt{``cat''} & \texttt{`a'} (character) \\
Lower or upper case? & \texttt{``APPLE''} & \texttt{`a'} (lower) \\
Vowel or character order? & \texttt{``pear''} & \texttt{`a'} (vowel order) \\
Special character if none? & \texttt{``Mmmm''} & \texttt{`-'} (hyphen) \\
\bottomrule
\end{tabular}
\label{tab:ambiguities}
\end{table}

A ``Probeable Problem'', as defined by Pawagi and Kumar \cite{pawagi2024probeable}, consists of a deliberately ambiguous problem statement and a mechanism for seeking clarifications regarding expected behaviour.  A Probeable Problem can be implemented as a traditional code-writing task (thus automatically gradable) along with an oracle that clarifies the expected output for any given input, or `probe'.  Table~\ref{tab:ambiguities} shows an example of how such probes can be used to clarify the missing details if asked to \emph{``find the first vowel in a string''}.


The use of Probeable Problems has not been explored at scale in an introductory programming context. In this work, we introduce such problems in the weekly laboratory sessions of a large introductory programming course. These problems can easily be configured using a standard auto-grader as a pair of tasks, each of which provides clarifying feedback. \emph{Probes} allow students to submit their own test inputs and observe the expected outputs, whereas \emph{code evaluations} check students' code against instructor-supplied tests and report any failing inputs. By imposing a small penalty on the latter (which mimics the `expensive' \emph{clarify after coding} behavior), we hope to encourage the `cheaper' and more desirable \emph{clarify before coding} behavior. We collect over 40,000 probes
and analyze how probing activity relates to task success. We also investigate student perceptions of these problems, in comparison to traditional programming exercises with well-defined requirements.  We organize our analysis around the following three research questions:


\begin{itemize} 
    \item \textbf{RQ1:} How frequently do students submit probes \emph{prior} to writing any code, and how does this relate to overall course performance?
    \item \textbf{RQ2:} To what extent do students rely on `cheap' (clarify before coding) probes compared with ‘expensive’ (clarify after coding) code evaluations? 

    \item \textbf{RQ3:} How do students describe their experiences with Probeable Problems compared to traditional programming tasks?

\end{itemize}


\section{Related Work}

\subsection{Learning to Code Without Ambiguity}

The idea of incorporating ambiguity into programming assignments dates back decades. In 1978, Schneider \cite{schneider1978introductory} argued that a key skill for CS1 students is the ability to ``recognize and resolve uncertainties in simple problem statements''.  At the time, he proposed that ``some programming assignments should intentionally be left incomplete, requiring the student to consider the alternatives and to make a reasonable decision on the omitted details''. Despite its promise, this approach has seen limited adoption in introductory programming education. Instead, the widespread popularity of automated assessment tools has reinforced a focus on solving well-defined problems with clearly specified, unambiguous requirements \cite{paiva2022automated, messer2024automated}. This exercise-intensive approach remains central to introductory programming pedagogy, as highlighted by Luxton-Reilly et al. in their comprehensive review of the literature \cite{luxtonreilly2018introductory}.

While traditional code-writing exercises are effective for developing mastery of syntax and basic programming concepts, they tend to neglect higher-order cognitive and metacognitive skills such as handling ambiguous requirements, which is essential for real-world programming \cite{shah2015resolving, gervasi2019ambiguity}.  Moreover, the ubiquity of large language models (LLMs) presents an immediate challenge to syntax-focused pedagogies \cite{denny2024cacm}.  Modern LLMs are capable of solving well-defined programming tasks with very high accuracy \cite{savelka2023thrilled, gutierrez2024seeing, finnieansley2023myai, kiesler2023large}.  This has led to calls from the computing education community to explore new types of learning activities that are more authentic and relevant in the era of AI-assisted programming \cite{raman2022programming, franklin2025elements}.


\begin{table*}[htb]
\small
\caption{The Probeable Problem statements, the default probe provided to students (which does not reveal an ambiguity), and a brief summary of the ambiguities that students must identify to solve the problem. Model solutions (commented with ambiguities and their resolutions) are available at: \url{https://onlinegdb.com/8ny78ivwV}}
\label{tab:ProblemList}
\begin{tabular}{c p{6cm} l p{4.9cm}}
{\bf Problem} & {\bf Statement} & {\bf Default Probe} & {\bf Summary of ambiguities} \\
\toprule
P7 & Implement a function to count the number of integers between a and b in an array of length n & CountBetween(\{1, 2, 3\}, 3, 0, 5) $\rightarrow$ 3 & \emph{`a' and `b' can be specified in either order, and they are strictly excluded from the range} \\
P8 & Implement a function to search an array of length n for the smallest even value & SmallestEven(\{50, 25, 2, 30, 45\}, 5) $\rightarrow$ 2 & \emph{multiple indices are printed in descending order, and special error printed if no evens} \\
P9 & Implement a function to find the first vowel in a string & FirstVowel(``apple'') $\rightarrow$ `a' & \emph{details in Table \ref{tab:ambiguities}; return lower case character in vowel order, special character if no vowel} \\
\bottomrule
\end{tabular}
\end{table*}

\subsection{Embracing Ambiguity}

While problem statements in introductory courses are usually well-specified, this does not guarantee that students develop a correct understanding of the problem before attempting to solve it. One approach to address this challenge is to have students create and solve test cases. For example, Craig et al. \cite{craig2019answering} and Denny et al. \cite{denny2019closer} demonstrated that having students solve or create test cases before writing code can enhance their understanding of problems and reduce errors, even when a problem is well specified.  Similarly, Pechorina et al. designed a metacognitive scaffolding tool which encouraged students to write and solve test cases prior to implementing any code, which reduced errors during implementation \cite{pechorina2023metacodenition}.  
This confirms earlier findings from Prather et al. showing that 
when students solve a test case after reading a problem statement, they build more accurate mental models of the problem  \cite{prather2019first}.
Wrenn and Krishnamurthi highlighted that flawed problem comprehension can lead to flawed implementations, 
particularly when students develop tests and code based on the same misunderstanding \cite{wrenn2019executable}.  
They addressed this by introducing Examplar, which provides feedback on input-output examples to validate problem comprehension.  


Unlike Examplar, Pawagi and Kumar directly focus on the issue of ambiguous requirements when introducing the idea of Probeable Problems \cite{pawagi2024probeable}, where essential details are deliberately omitted from the problem statement. They evaluated this idea in a two-hour programming contest with 67 undergraduate 
students from multiple institutions. While students could often identify some of the missing details through probing, they frequently missed others, especially subtle omissions like tie-break mechanisms. Although students 
could use AI tools like GitHub Copilot or ChatGPT, the incomplete specifications meant these tools did not guarantee correct solutions.

In the current work, we apply the concept of Probeable Problems for the first time to a large introductory programming course and focus on analyzing both student perceptions and probing behaviors at scale.  We believe that Probeable Problems, which can be implemented using existing automated grading tools, offer an exciting way to build on Schneider's 50-year old vision.  They 
require students to uncover missing specification details through an iterative process of writing test inputs and obtaining oracle feedback, addressing a skill often neglected in traditional CS1 pedagogy.

\section{Methods}
Our exploration of Probeable Problems takes place in the context of a large introductory programming course taught at the University of Auckland in the second half of 2024, with a total enrolment of 1028 students.  This first-year course is compulsory for all engineering students, who only select specializations in their second year.   We identified three consecutive labs in the weekly lab schedule, and introduced one Probeable Problem into each lab alongside several traditional programming exercises.  For convenience, we will refer to the three Probeable Problems as \pr{7}, \pr{8}, and \pr{9}, where the numbers correspond to the week of the course.  Table~\ref{tab:ProblemList} lists the problem statements and the default probe (provided to students) for each.  
Model solutions (commented with ambiguities and their resolutions) are available online\footnote{\label{sharednote}\url{https://onlinegdb.com/8ny78ivwV}}.

\begin{figure}[h]
  \centering
  \includegraphics[width=\linewidth]{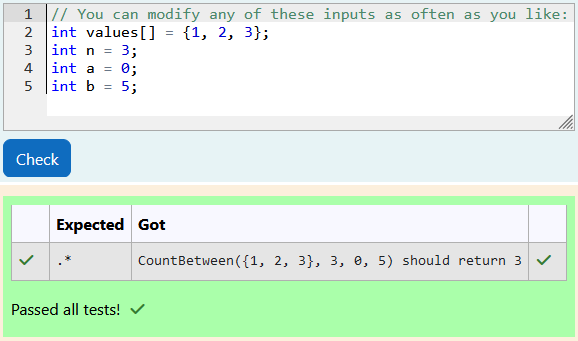}
  \caption{The CodeRunner platform, a widely used automated grading tool, configured to accept probes for P7: \emph{``Implement a function to count the number of integers between a and b in an array of length n''}. Students modify inputs in the upper editing pane, and view the expected output in the lower pane.}
  \Description{CodeRunner screenshot}
\label{fig:CodeRunner}
\end{figure}

The course uses the CodeRunner platform for automating the grading of programming exercises \cite{lobb2016coderunner}.  
Instructors set up programming exercises by providing a problem statement, a set of test case inputs with corresponding expected outputs, and a model code solution (which is optional, but allows CodeRunner to verify the test cases are valid during problem authoring).   CodeRunner can be configured to use regular expressions when matching expected outputs, and so the `probe' component of a Probeable Problem can be configured using a single test case with a wildcard to successfully match any output.  To serve as an oracle, the template simply includes a model solution for the problem (not shown to students) which prints the output produced when called using the student-provided inputs.  
Figure \ref{fig:CodeRunner} illustrates the interface for accepting probes.


Consistent with course policy, for programming exercises a small grading penalty applies for incorrect submissions (i.e. code submissions that do not completely pass the test suite).  
A penalty is applied to the final score for a given problem, which grows in 5\% point increments for multiple incorrect code submissions.  This is a small penalty relative to the course -- each lab contributes 1\% towards the final grade, and a single programming exercise contributes no more than 10\% to a given lab.  
Each Probeable Problem consists of the probe component (shown in Figure \ref{fig:CodeRunner}) and the corresponding programming exercise.  No penalties were applied for submitting probes -- students could submit as many probes as they wished -- but the standard penalties applied for incorrect code submissions. 

In terms of describing these tasks to students, we followed the approach of Pawagi and Kumar and referred to them as ``Ask the Client'' questions.  The framing shown to students at the start of each question was as follows:

\begin{quote}    
    \emph{This is a ``Ask The Client'' question.  You have been approached by a mysterious client who would like you to write a program for them, however, they have been a little vague. They have asked you to...}
\end{quote}

This framing was then followed by the problem statement, and a reminder that no penalty was associated with submitting probes.

\subsection{Quantitative Analysis}
We define an \emph{attempt} as the complete sequence of $P$'s (probes), $F$'s (failing code submissions) and $S$'s (successful code submissions) submitted by an individual student. Excluding the attempt for one student who dropped the course after the \pr{7} assessment, we logged 2,896 attempts across all three problems, containing a total of 44,068 probes. Of these, 3,649 probes were the defaults we had provided. As shown in Table~\ref{tab:ProblemList}, each problem's default probe was chosen to avoid revealing any ambiguities about that problem. For our analysis, we eliminated these default probes. For example, the attempt $PPFS$ represents two non-default probes, followed by a failing submission, and ending with a successful submission.

\begin{figure*}[h]
  \centering
  \includegraphics[width=\linewidth]{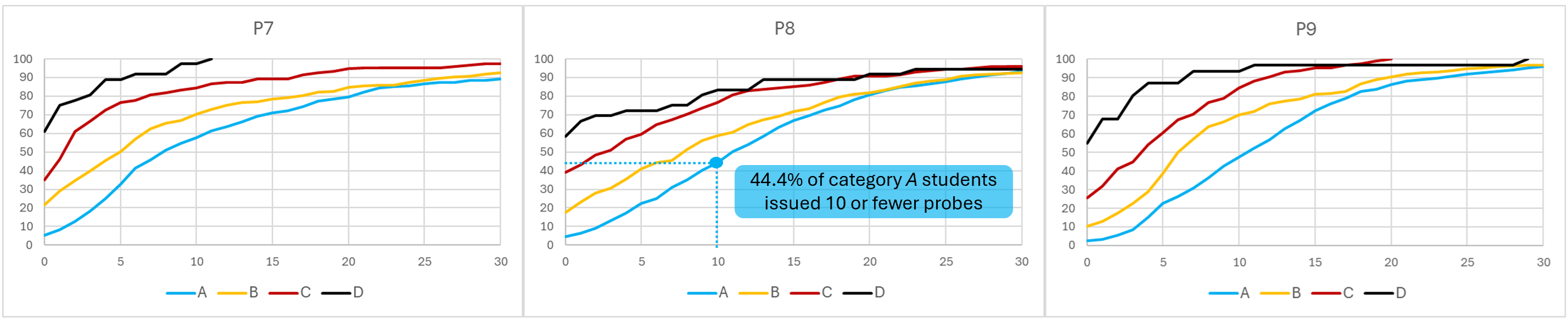}
  \caption{The cumulative percentage of students ($y$-axis) in each grade category for each Probeable Problem as a function of the number of probes ($x$-axis) submitted \emph{prior to} the first code submission ($F$ or $S$).}
  \Description{For each problem, when the number of probes is zero, weaker students (with poorer grades) start with a higher cumulative percentage than those with better grades. Also, the curve for weaker students generally approaches 100\% faster than for better students. Thus, weaker students issue substantially fewer probes.}
\label{fig:q_before_code}
\end{figure*}

\begin{figure*}[h]
    \centering
\includegraphics[width=\linewidth]{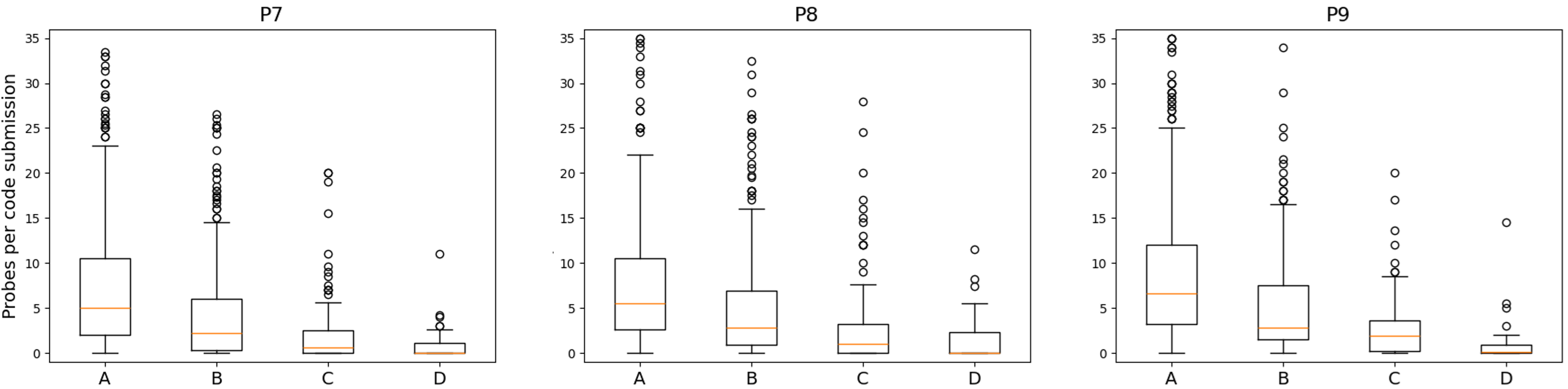}
    \Description{The boxplots show two major trends. First, for all three problems, both the variance and the median decrease as we move from category $A$ to $B$ to $C$ to $D$. Second, the median value increases slightly for category $A$ students, but remains close to zero for category $D$ students.}
    \caption{Box plots showing the \emph{ratio} of probes issued per code submission in each attempt, across all grade categories and problems. We exclude outliers where this ratio exceeds 35 (i.e., 2.9\%, 1.3\%, and 0.7\% of \pr{7}, \pr{8}, and \pr{9} attempts respectively).}
    \label{fig:probe_failures}
\end{figure*}

Fewer than 2\% of attempts were just $S$ i.e., a single successful attempt without \emph{any} probes. We find it implausible that these were honest attempts. Hence, we reject these attempts from our subsequent analysis in this paper (although note that future work could explore such attempts as an indicator of academic misconduct). In addition, we excluded attempts that did not include at least one code submission ($F$ or $S$). We focus on the remaining 930 attempts for \pr{7}, 925 attempts for \pr{8}, and 901 attempts for \pr{9}. The vast majority of these attempts were successful (i.e., they included $S$). The proportion of all attempts that were successful 
was considerably higher for \pr{7} (878/930) than for \pr{8} (809/925) and \pr{9} (801/901). 

\subsection{Thematic Analysis}

One member of the research team conducted a reflexive thematic analysis~\cite{clarke2017thematic} to understand how students perceived this novel problem type. We asked students to respond to the question: \emph{``Please comment on your experience solving this type of `Ask the client' task (where you must create input tests to clarify the behavior of a vague problem statement)
compared to a more traditional programming task where a complete and detailed specification is provided to you.''} 

Students' responses were first open-coded by a single researcher. Next, the codes were iteratively reviewed and organized into themes based on their relevance to RQ3. Finally, to help contextualize the prevalence of the themes, we deductively counted the number of instances for each theme. This final step is not a necessary component of thematic analysis, but given the number of participants, it provided helpful context for understanding the themes. 

\section{Results}
For our analysis, we partition attempts into four categories based on the eventual course grade received by the student that made the attempt: category $A$ (grades A+, A, or A-), category $B$ (grades B+, B, or B-), category $C$ (grades C+, C, or C-), and category $D$ (grades D+, D, or D-).  Category $D$ grades are failing grades in the course. 

\subsection{RQ1: Probing Before Coding}
Prior to the introduction of Probeable Problems on Lab~7, students were accustomed to immediately start writing code. Since we encouraged students to probe \emph{before} coding, we hoped to see attempts starting with several $P$'s before the first code submission ($F$ or $S$). Figure~\ref{fig:q_before_code} shows the \emph{cumulative} percentage of students for each grade category as a function of the number of probes before their \emph{first} code submission ($F$ or $S$). For a given number of probes $p$ on the $x$-axis, observe that the cumulative percentage of students ($y$-axis) generally decreases as we move from category $D$ to $C$ to $B$ 
to $A$. For instance, on all three problems, more than half the students in category $D$ issued \emph{no} probes before starting to code. In contrast, only 5.1\% of category $A$ students started coding without probing on \pr{7}, and this percentage fell to 2.4\% on \pr{9}. We find it encouraging that students in all categories appeared to see value in probing as they gained familiarity with Probeable Problems. Thus, while fewer than 25\% of students from category $C$ issued more than 5 probes before coding on \pr{7}, this proportion increased to about 40\% on later problems. Interestingly, more than 10\% of students in category $A$ issued over 30~probes when first exposed to Probeable Problems (\pr{7}), but this percentage fell to less than 6\% on the remaining problems. 



\subsection{RQ2: `Cheap' vs. `Expensive' Clarifications}
We encourage students to discover and clarify ambiguities via `cheap' probing by imposing no penalty on probes. Students who fail to discover certain ambiguities may submit code that resolves these ambiguities differently from our model solution. Such code is likely to fail at least one test case, and we show students the first such failing test case to help them recognize this ambiguity. This is a more `expensive' way to discover and clarify ambiguities, since we impose a small penalty for failing code submissions.


Figure~\ref{fig:probe_failures} shows the ratio of `cheap' probes versus `expensive' code submissions issued by students. For each problem, the median of this ratio decreases from category $A$ to $B$ to $C$ to $D$, suggesting a declining ability to imagine ambiguities (and thus a relatively higher reliance on `expensive' clarify after coding feedback). While this is concerning, suitable  interventions supplemented with many Probeable Problems may help all students develop this ability.


\subsection{RQ3: Qualitative Results}

Our thematic analysis resulted in six key themes: difficulty and frustration, real-world relevance, learning benefits, preference for traditional tasks, engagement, and problem-solving strategies. 

\subsubsection{Difficulty and Frustration}

The most common theme (436 responses) was that students found the problems to be difficult, and at times, frustrating. The vague problem descriptions and hidden requirements were challenging, and students reported feeling overwhelmed by the need to identify edge cases and account for unanticipated scenarios. One student said, \textit{``I find these tasks often quite difficult because of how vague the question is. This means that I have to think ahead of time to determine different variables and how each of the sections of code will be laid out.''} Similarly, another student shared, \textit{``I struggled a lot when trying to find what the client is asking for. Once I found out what was being asked, I solved it, but there were things I forgot to check such as what if there is no evens.''}

\subsubsection{Real-World Relevance}

Students (348 responses) reported appreciating the alignment between these tasks and real-world programming scenarios. Many recognized that working with vague specifications and clarifying requirements are essential skills for developers who interact with clients in professional contexts. One student commented, \textit{``I enjoyed it; I thought it was more accurate to real-life examples where the client is not a coder and cannot describe the code that well but can give examples of exactly what they like.''} Another wrote, \textit{``This type of exercise mimics a situation where a client asks the programmer to create a program in a vague manner. It makes me think consciously through every possible scenario and what the output of the program should be.''} Despite the prevalence of this theme, some students also expressed unrealistic expectations about requirement elicitation. For example, one student said, \textit{``In reality, the client should provide a full list of requirements of what to achieve, or they should provide more details on a specific component.''} Another student said, \textit{``I hate clients that aren't clear about what they want!''} This highlights a disconnect for some students who may not yet view requirement elicitation as an important skill.



\subsubsection{Learning Benefits}

Many responses (242) described the learning benefits that students believed these tasks provided. For example, the process of probing for requirements and addressing edge cases improved their understanding of programming concepts and problem-solving strategies. A student described this by saying, \textit{``These tasks helped me think more deeply about the problem and discover aspects of the specification instead of just having it provided to me.''} Another said, \textit{``It was frustrating at times, but it helped me improve my problem-solving skills and learn how to communicate better in code.''} In addition to improving problem-solving skills, students described needing to think creatively. For example, a student said, \textit{``It creates a fun and more problem-solving style approach to the questions. It also allows me to think outside the box and challenge how I interpret the results.''} Some students also described metacognitive benefits such as planning, reflection, and strategy development. For instance, a student shared, \textit{```Ask the Client' tasks challenge me to clarify vague requirements that reflect real-world scenarios. They improve problem-solving and communication skills by focusing on understanding the problem before coding, unlike traditional tasks with clear instructions.''} Although it is not yet clear whether these benefits will translate to traditional programming tasks, it is encouraging that so many students described these metacognitive benefits in their responses.

\subsubsection{Preference for Traditional Tasks}

Most of the students (312 responses) preferred traditional programming tasks with detailed problem specifications. They described well-specified tasks as being more straightforward and less time-consuming. One student stated, \textit{``I much prefer traditional programming tasks where it is less guessing what is wanted. With the `Ask the Client' tasks, it is very easy to forget certain edge cases, which can be frustrating.''} Students often referenced the time wasted by probing for requirements. One student shared, \textit{``It can be a bit of a pain due to the lack of information provided from the client...[which] can lead to going down many rabbit holes and causing time to be wasted.''} 

\subsubsection{Enjoyment or Engagement}

Despite the challenges, many students (247 responses) described enjoying or feeling engaged by the experience of uncovering requirements and edge cases. The challenge was intellectually stimulating. One student noted, \textit{``I enjoyed these tasks; they are a good way to remember to make sure I am testing for edge cases and fully understand what the code is supposed to do.''} Another said, \textit{``I really like these questions! They challenge me to think critically and improve my ability to look for edge cases.''} 




\subsubsection{Strategies and Problem-Solving}

Finally, 198 responses highlighted the strategies students developed to tackle the `Ask the Client' tasks. Many shared their approaches for identifying edge cases, systematically testing inputs, and iterating on their solutions. One student explained, \textit{``I found it quite important to begin by trying all the potential input cases to see what the expected output should be so I got an idea of how to handle boundary cases and typical inputs.''} Another said, \textit{``I learned to be more critical in finding possible edge cases to test. In future tasks, I will brainstorm edge cases before starting to code.''} 

Overall, the thematic analysis shows that while the task is difficult and can be frustrating, many students described experiencing benefits related to learning or metacognition. They appreciated the real-world relevance of needing to clarify problem specifications and found the challenge to be intellectualy stimulating.

\section{Discussion}

One interesting aspect of the probing behavior is that it seems to encourage metacognitive reflection. Metacognition, which is often termed ``thinking about thinking'' is a crucial skill in programming problem solving \cite{prather2020what}. Recent work by Bubnic et al. has shown that student metacognition can predict success in programming problem solving and could be one of the most important predictors of success in introductory programming \cite{bubnic2024metacognition}. Prather et al. have noted introductory programming students often struggle with metacognitive difficulties, such as solving the wrong problem and a lack of awareness as to where they are in the problem solving process \cite{prather2018metacognitive}. The results of the thematic analysis above showed that students were made aware of their own metacognition in ways that combat these issues, such as thinking more critically and deeply about the problem before beginning and that identifying the boundaries of the problem encourages finding edge cases. While metacognition was not directly measured in this experiment, it is often difficult to measure and student reflections can serve as a proxy \cite{loksa2022metacognition}. Furthermore, recent work has noted that student metacognition and self-regulation suffer dramatically when solving traditional programming assignments with LLMs \cite{prather2024wideninggap}, and can decrease critical thinking skills \cite{jost2024impact}. Therefore, Probeable Problems seem to be precisely the kind of metacognitive scaffolding 
that the computing education research community has been calling for 
in response to the threats of LLMs \cite{denny2024cacm, prather2023robots, prather2024wideninggap}.

Recent work on novice programmer use of generative AI reveals several striking similarities to our findings presented above. Margulieux et al. reported that higher performing students tended to use AI less and later in the problem solving process \cite{margulieux2024selfreg}. This shows a tendency by high performers to try and solve issues themselves before seeking help from AI tools. 
Our results above showed that students in the \textit{A} category submitted more probes before submitting code (see Figure \ref{fig:q_before_code}), attempting to find the edge cases themselves before hitting one with a failed code submission. These differences in behavior between grade categories may be interpreted through the lens of Expectancy-Value Theory \cite{wigfield2000evt}, which suggests that students are more likely to engage in a task when they both value it and believe they can succeed.
Students who recognized the real-world relevance of Probeable Problems, which was a prominent theme in our qualitative analysis, may have seen them as more worthwhile and thus engaged more deeply with the probing process.

Encouragingly, we observed a positive trend in all groups over time as they progressed from P7 to P9, with more students submitting probes before coding and more probe submissions overall. Moreover, the thematic analysis revealed that Probeable Problems encouraged students to engage in thorough planning. This suggests that Probeable Problems may provide a useful way to scaffold metacognition in novices, and promote critical thinking about programming that could decrease over-reliance on generative AI. 

\section{Future Work}
While requirement elicitation is often reserved for upper-level courses, such as Software Design or Project Based Capstone Courses \cite{morkos2019investigating, kumar2024computer}, the ability of LLMs to solve well-specified problems creates an opportunity to integrate this critical skill earlier in the curriculum via Probeable Problems. 
Future work could explore how early exposure to Probeable Problems supports students in the lower-performing categories, particularly those at risk of failure ($D$ category), by scaffolding critical skills like metacognition and problem-solving. These skills are not only essential for navigating less well-specified labs or assignments in subsequent courses but are also vital for tackling real-world programming challenges, where ambiguity is a common obstacle.

Future work could also explore alternative grading strategies.  While the current approach does not penalize probing, it may be possible to incentivize thoughtful engagement with the probing process. For example, penalties could be introduced for making many redundant probes, or the code implementation step could be disabled until a sufficient number of probes have been submitted to clarify the ambiguities in the problem. Experimenting with a variety of grading models that emphasize more strategic probing could deepen students' critical thinking and task engagement. 


Finally, the ``Ask the Client'' framing could be further enriched through natural language or multimodal interfaces that simulate real-world client interactions. Generative AI agents could simulate the client, allowing students to ask clarifying questions in natural language or engage through voice. This would create a more realistic and immersive experience, helping students build both technical and communication skills essential for professional work.

\section{Limitations}


One limitation of this work is that the analysis focuses on the quantity of probes submitted by students. While this provides useful insights into engagement with Probeable Problems, it does not account for the quality or types of test cases used. A richer analysis that examines the diversity, relevance, and depth of submitted probes could yield a better understanding of how students approach problem clarification.  Also, it is possible that motivated students who are performing well in the course are more likely to engage in systematic probing. The observed correlation between probing behavior and academic success does not establish causation and warrants further investigation.



\section{Conclusion}
We explored the use of Probeable Problems~\cite{pawagi2024probeable} in an introductory programming course, 
where students clarified vague problem specifications through probing.
Our results show that students appreciated the real-world relevance of these tasks and reported benefits such as improved problem understanding, greater attention to edge cases, and enhanced problem-solving strategies. Higher-performing students engaged more systematically with probing, submitting more test inputs before coding and relying less on failed code submissions to clarify ambiguities.  While some students found the ambiguity frustrating, many described the tasks as intellectually stimulating and reflective of real programming practice. Probeable Problems offer a scalable way to foster critical thinking and metacognitive skills, and may help address some of the challenges posed by generative AI in programming education.







\begin{acks}
This research was supported by the Research Council of Finland (Academy Research Fellow grant number 356114).
\end{acks}

\bibliographystyle{ACM-Reference-Format}
\bibliography{references}

\end{document}